\documentclass[a4paper,11pt]{article}
\usepackage{jinstpub}
\usepackage{subcaption} %to split figures with subcaptions
\usepackage{multicol} %to split in columns
\usepackage{url} %to include links
\usepackage{hyperref} %to reference formulas..
\usepackage{hepnames} %for particles
\usepackage[sicmds,freestanding]{hepunits} 
\usepackage{booktabs} %for cool table lines 
\usepackage{chemformula} %for chemical components (eg., \ch{^{90}Sr}, \ch{SF6})

\usepackage[toc,page]{appendix}
\title{Measurement of Single Event Effect cross section induced by
  monoenergetic protons on a $130$~nm ASIC}
\author{D. Boumediene, F. Jouve, D. Lambert, R. Madar, S. Manen, O. Perrin, L. Royer, A. Soulier, R. Vandaele}
\affiliation{Laboratoire de Physique de Clermont-Ferrand (LPC), Universite Clermont Auvergne, CNRS/IN2P3,\\4 avenue Blaise
  Pascal, 63178 Aubi\`ere, France}
\emailAdd{djamel.boumediene@cern.ch}

\abstract{ %%
  Designing integrated circuits in radiation environments such as the High Luminosity LHC (HL-LHC) is challenging.
  Integrated circuits will be exposed to radiation-induced Single Event Effects (SEE).
  In deep sub-micron technology devices, the impact of SEEs can be mitigated by triple modular redundancy.
  The triplication of the most sensitive data is used to recover most of the data corruption induced by interacting particles.
  One type of SEE, called single event upset (SEU), is studied in this paper. The SEU cross-section and the
  performance of the triplication are estimated using an ASIC prototype exposed to a beam of protons.
  The SEU cross-section is measured and systematic difference between $1\rightarrow{0}$ and $0\rightarrow{1}$ bit flip rates is observed. 
  The efficiency of the mitigation method is investigated.
}


\begin{document}
%\linenumbers
\maketitle

%% --------------------------------------
\section{Introduction}

Designing integrated circuits in radiation environments such as the High Luminosity LHC (HL-LHC) is challenging.
Integrated circuits will be exposed to radiation-induced Single Event Effects (SEE), especially Single Event Upsets (SEU).
In deep sub-micron technology devices, the impact of SEEs can be mitigated by triple modular redundancy~\cite{tmr,tmr2}.
The triplication of the most sensitive data is used to recover the data corruption induced by interacting particles.

SEU can be studied with different types of ionizing particles~\cite{method}. Heavy Ions can be used, offering the advantage of testing an extensive range of Linear Deposit Energies (LETs). However, a model has to be used to extrapolate the measurements to the final environment. Protons of at least $20~\MeV$ can be used to estimate the SEU effect, especially for the HL-LHC environment where the detector components are shielded against low-energy particles and where the expected fluences are usually expressed in unit number of $1\,\MeV$ neutrons equivalent per squared centimeter, $\mathrm{n}_{\rm{eq}}^{1\,\MeV} \centi{\meter}^{-2}$ unit. The energy dependence can be taken into account to refine the predictions.

In this paper, the SEU cross-section and the performance of the triplication are estimated using an ASIC prototype exposed to a monoenergetic beam of protons. This tested ASIC mainly contains a static random-access memory (SRAM) system made of flip-flops, capable of handling a latency of $38.4\,\mu$s, required by the trigger system of the experiment. This SRAM is naturally sensitive to SEU, mitigated by auto-correction using triplication.

This paper aims to determine the bit flip rate on a 130~nm technology. The corresponding SEU cross-section $\sigma_{\rm{SEU}}$ was measured in~\cite{dca}
and is expected to be in the order of $\sigma_{\rm{SEU}} = (6 \pm 1)\,10^{-14} \centi{\meter}^{2}$.
This rate, combined with a triplication design, leads to a residual autocorrection failure rate. This failure rate is estimated for
an 1600 bits ASIC, and it has to be negligible during an average data-taking period of 1 day.

Section~2 describes the facility and the setup used to collect the data. The data are analyzed in Section~3, and three results are presented:
\begin{itemize}
\item the SEU cross-section for single bits,
\item an estimate of the triplication failure rate using as a study case a 1600 triplicated bit ASIC at the HL-LHC, and
\item the direct measurement of triplication failures.
\end{itemize}
 %% --
%% --------------------------------------

%% --------------------------------------
\section{Experimental setup}
\subsection{The MEDICYC facility and beam monitoring}

The SEU cross-section was measured during a test beam campaign at the Centre Antoine Lacassagne (CAL) in 2021, where the MEDICYC cyclotron~\cite{medicyc,medicyc2} provides proton beams.
MEDICYC is equipped with several proton beamlines, two of which are in operation at the moment. A clinical beamline provides proton therapy of ocular melanoma. A research beamline has recently been equipped and is proposed to researchers and industry for R\&D programs.

 The subsequent measurements were performed using the low energy
treatment line, where a $62.0\,\MeV$ proton beam was produced with up to $300\,\nano{\ampere}$ intensity. This energy corresponds to a LET
 value of $8.7\,\keV\,\centi{\meter}^2/\milli{\gram}$ in silicon dioxide.
The beamline calibration and monitoring control the intensity at the 3 per mil level at the highest intensity. 

\subsection{Integrated circuit and data taking}

The $130\,\nano{\meter}$ TSMC ASIC prototype was designed to test the performances of digital blocks when exposed to radiation. One of these tests is dedicated to the SEE study. Sixty-four 8-bit registers were implemented and accessible through an $\mathrm{I^{2}C}$ port. They are fully dedicated to SEU counting, and their outputs have no other function in the chip. Each bit of a register is triplicated, and automatic corrections are applied to every triplicated-bits of the register at $40~\mega{\Hz}$ using a majority voter logic.  

\begin{figure}[t]
 \begin{center}
    \includegraphics[width=0.5\textwidth]{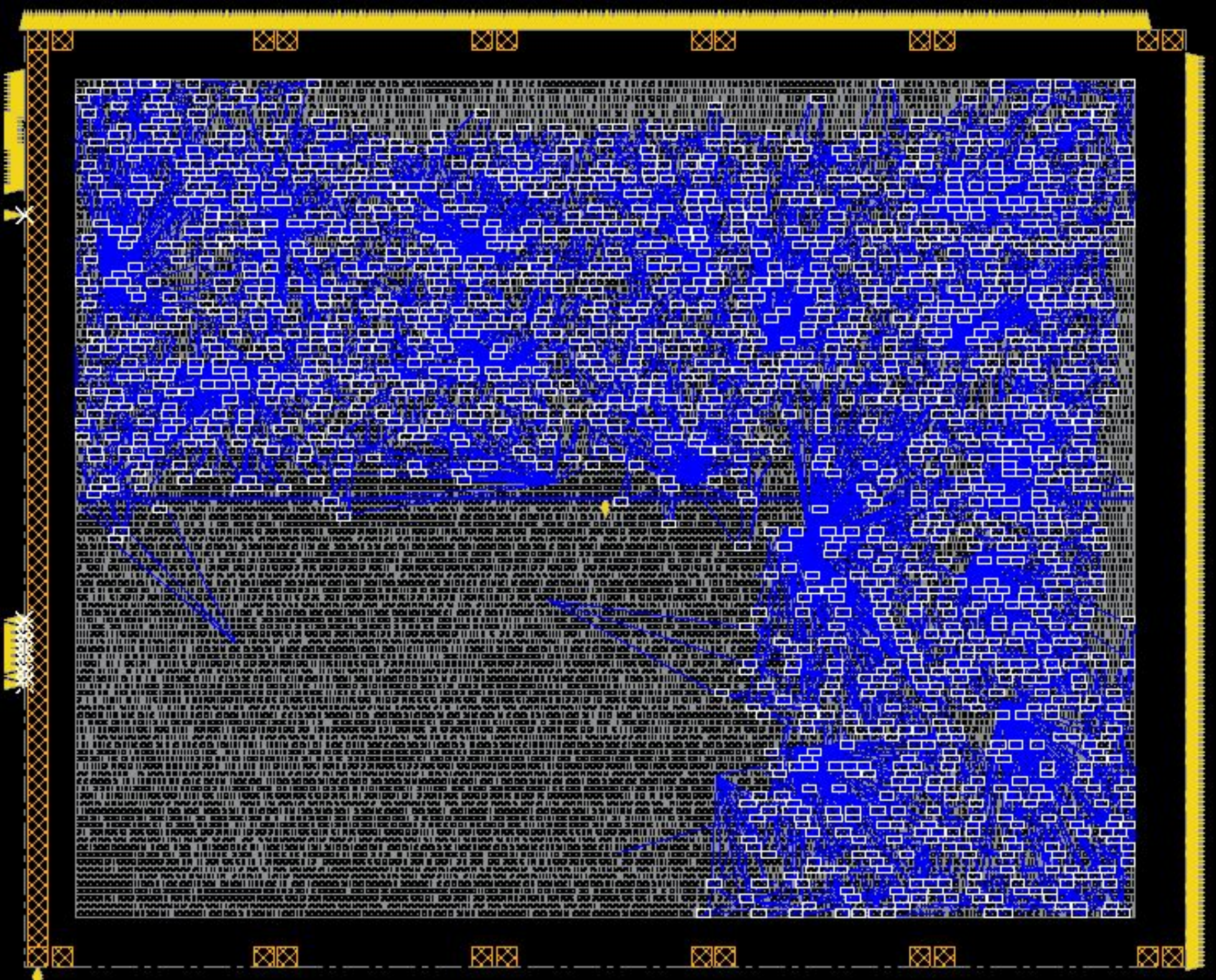}
  \caption{View of the ASIC where the flip-flops are highlighted by white rectangles. \label{fig:flipflops}}
  \end{center}
\end{figure}

\begin{figure}[t]
 \begin{center}
    \includegraphics[width=0.5\textwidth]{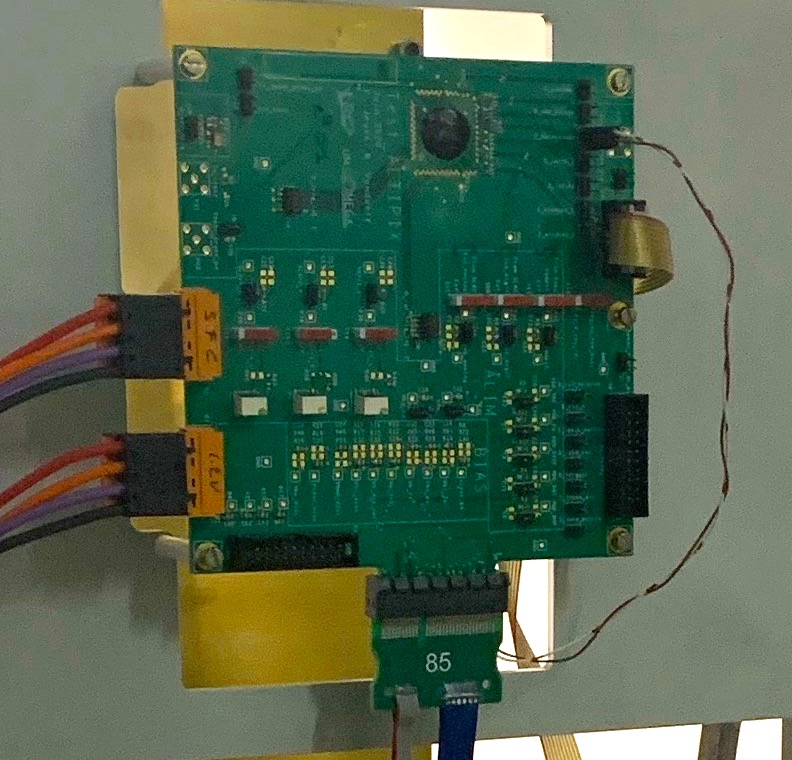}
  \caption{View of the ASIC (located under a black resin) on its host board mounted on the beamline.\label{fig:altipix}}
  \end{center}
\end{figure}

Each triplicated flip-flops are physically spaced by $15~\micro{\meter}$ in the layout and shown in Fig.~\ref{fig:flipflops}. The implementation of the Register Transfer Level (RTL) is done with a 2008 version of the CERN TMRG tool~\cite{tmrg}. It is integrated with the digital flow to automatize the process of triplication. The TMRG toolset generates constraints that ensure that triplicated items are not placed too close to each other while allowing the designer to decide which blocks or signals will be triplicated.
The clock, reset, and control signals are not triplicated in the ASIC prototype. The reset is synchronous with the $40~\mega{\Hz}$ clock, and each register can have a different reset value.  
If only one flip-flop has its value changed due to SEU, the
voted value is overwritten inside the faulty flip-flop thanks to the
auto-correction system. The voter also raises an error flag if the three flip-flops do not have the same value. This error flag is then used to increment counters to count SEU.
Four counters are associated with the correction mechanism. Each counter covers 16 registers, called groups, and is incremented by one if a correction is applied to any one of them. The four counter values are stored in triplicated registers. 
The total number of bits monitored in the SEU rate measurement is 1536.
The ASIC is mounted on a custom board, shown in Fig.~\ref{fig:altipix}, that routes the power supplies, the clock and various probe signals.

A microcontroller, linked to an $\mathrm{I^{2}C}$, is used to communicate with the ASIC, read or reset the counters, and control the register contents. External devices ensure a 40~MHz clock as well as a power supply. 
The various counters were readout, and the registers reinitialized, after each reading, with predefined words at a typical frequency of $100~\Hz$. 

The ASIC was aligned with the proton beam covering the device surface with a homogeneity better than 5\%.
Data were recorded using two beam intensities during several data-taking periods, four at high intensity and one at low intensity.
The ASIC was exposed to a total fluence of $8.7~10^{14}~\mathrm{p}\,{\centi\meter}^{-2}$.

 %% --
%% --------------------------------------

%% --------------------------------------
\section{SEU Cross-section measurement and auto-correction failure}

Two methods are presented to measure the SEU cross-section. The first one relies on the error rate measurement (section~\ref{subsec:xsec}).
The second method exploits the randomness of the SEU, which must result in duration between two consecutive errors $\Delta t$ distributed according to a decreasing exponential (section~\ref{subsec:dt}).

Two methods are presented to estimate the auto-correction failure rate. The first one is based on the probability of having two simultaneous SEUs in a triplicated system. It is computed from the measured SEU cross-section (section~\ref{subsec:computation}).
The second method consists in counting the actual number of events in which the auto-correction failed (section~\ref{sec:xsfail}). For completeness, a Monte Carlos simulation is performed and
compared to data (section~\ref{sec:sim}).

\subsection{SEU cross-section}
\label{subsec:xsec}

\begin{figure}[t]
  \includegraphics[width=\textwidth]{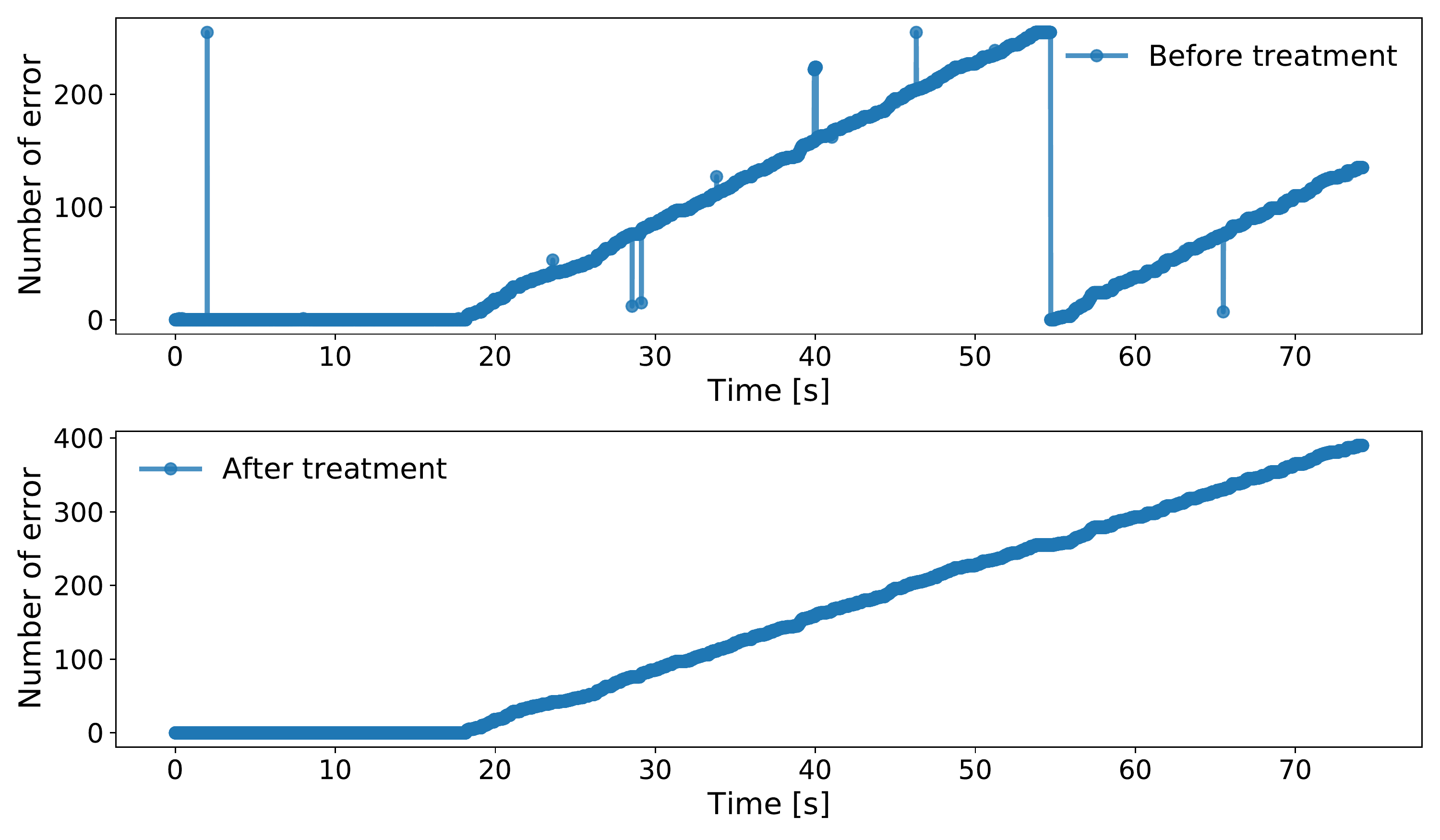}
  \caption{ Cumulative number of auto-corrections before (top)
    and after (bottom) offline corrections from data corruption
    and overflow in the counter. \label{fig:err}}
\end{figure}

\begin{figure}[t]
  \includegraphics[width=\textwidth]{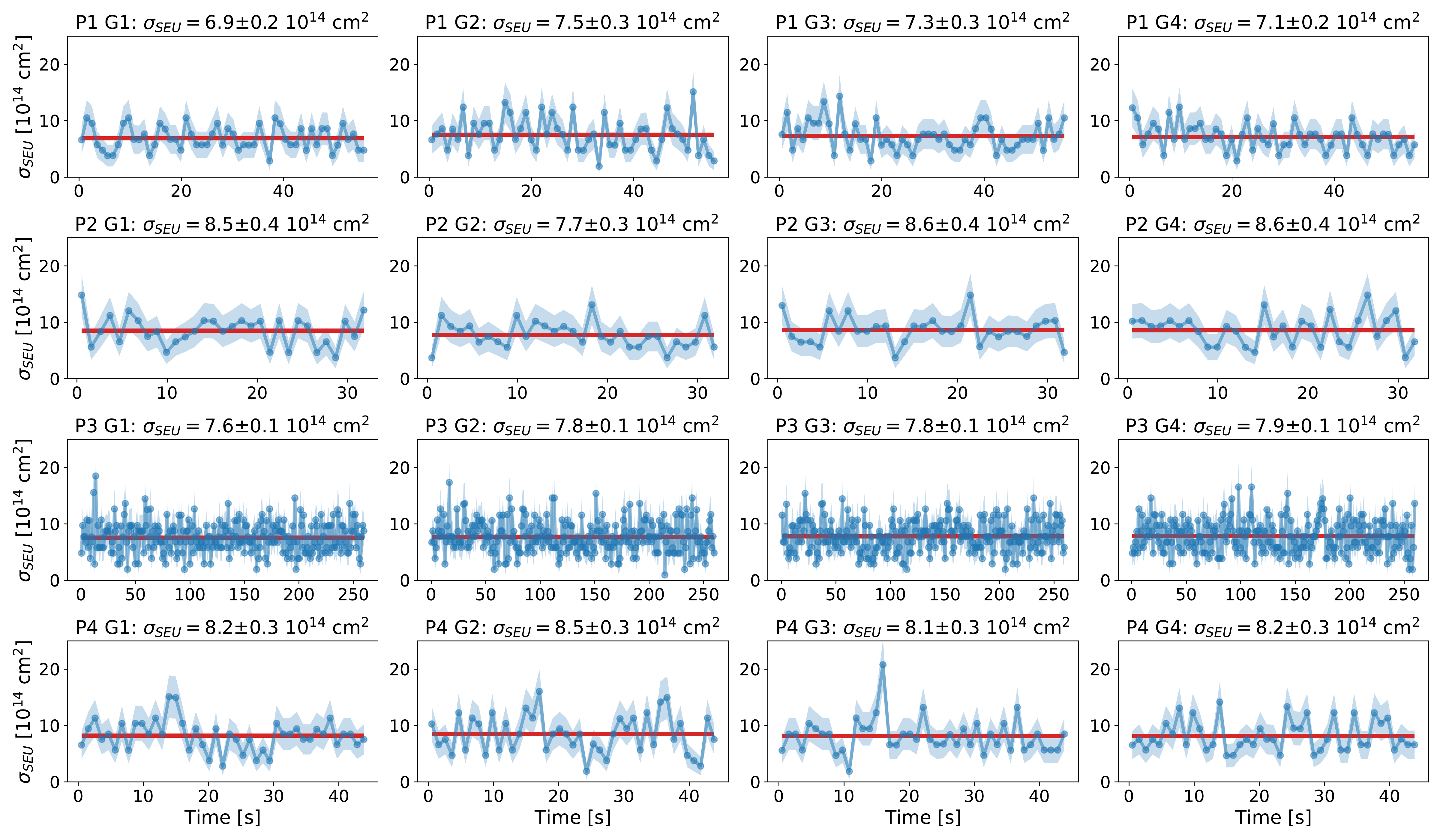}
  \caption{SEU cross-section computed every second for the four groups (row) of the four periods (column).
    The red line corresponds to the fit result over the whole period. The index of the group and period and the fitted value are specified on top of each sub-figure. \label{fig:slope_vs_time}}
\end{figure}

The registers were set to {\ttfamily{0}} during one period and to {\ttfamily{1}} during four periods.
The SEE rate estimate relies on the auto-correction counters.
These SEE counts are expected to be mainly induced by SEU.
A few jumps in the counters were observed and are interpreted as
non-corrected SEEs on the counter itself. These jumps are then filtered in order to keep only jumps related to SEU.
Also, when a counter reaches 255 it goes back to zero at the next incrementation. These resets are also corrected.
The number of errors before and after treatment is shown in Fig.~\ref{fig:err} as a function of time for one of the periods and groups.
The number of errors normalized to the beam intensity is used to extract the SEU cross-section per bit, as summarized in Table~\ref{tab:xs}.
A rate is computed for each period and group, and a global rate is obtained by fitting these individual values assuming an error occurrence following a Poisson distribution. The rate as a function of time, as well as the fit result for the sixteen datasets at high intensity, are shown in Fig.~\ref{fig:slope_vs_time}.
The final rate value, and its uncertainty is derived from the mean and the mean error over the different periods and groups.
 %The uncertainty covers beam inhomogeneities, beam intensity, and statistical errors.
The average cross-section is estimated to be 
$$\sigma_{\rm{SEU}^{1{\rightarrow}0}} = (7.90 \pm 0.13)~10^{-14}\,\centi{\meter}^{2}$$
$$\sigma_{\rm{SEU}^{0{\rightarrow}1}} = (6.77 \pm 0.41)~10^{-14}\,\centi{\meter}^{2}$$
The systematic difference between
$1\rightarrow{0}$ and $0\rightarrow{1}$ bit flip rates is significant and
already reported, eg.\, in ref.~\cite{dca}.
The dependency to the LET, and therefore to the beam energy, as shown in ref.~\cite{method,seu.vs.energy}, has
to be considered for extrapolations at different energy scales. 
%
%%%%%%%%%%%%%%%%%%%%%%%% Table %%%%%%%%%%%%%%%%%%%%%%%%%%%
\begin{table}[t]
  \centering
  \caption{
SEE cross-sections, $\sigma_\mathrm{SEU}$, per bit for all the registers and different periods, measured using two beam
intensities, $I$. The 64 registers are combined in four groups numbered from 1 to 4. The uncertainties assigned to the measurement from indivdual periods combines the statistical error with the uncertainty on the beam intensity. The uncertainty on the average is computed using the spread of the individual measurements.
\label{tab:xs}}
\begin{tabular}{clll}
\toprule
$I~[\nano\ampere{\centi{\meter}}^{-2}]$ & Period & Group & $\sigma_{\mathrm{SEU}}~[10^{-14}\centi{\meter}^{2}]$ \\ \midrule
$3.1\pm 0.1$ & period 1 & 1 & $7.97 \pm 1.13$ \\ %$8.27 \pm 1.19$ \\ 
 &  & 2 & $6.37 \pm 1.01$ \\% $7.23 \pm 1.11$ \\ 
 &  & 3 & $5.74 \pm 0.96$ \\% $5.86 \pm 1.00$ \\ 
 &  & 4 & $7.01 \pm 1.06$ \\ \cline{2-4} % $6.54 \pm 1.06$ \\ 
 & Average &  & $6.77 \pm 0.41$ \\ \midrule % $7.03 \pm 0.90$ \\ \midrule

$299.9\pm 1.3$ & period 1& 1 & $6.88 \pm 0.25$ \\ 
 &  & 2 & $7.50 \pm 0.25$ \\  
 &  & 3 & $7.30 \pm 0.25$ \\  
 &  & 4 & $7.11 \pm 0.25$ \\  

 & period 2 & 1 & $8.52 \pm 0.36$ \\
 &  & 2 & $7.74 \pm 0.34$ \\  
 &  & 3 & $8.64 \pm 0.36$ \\  
 &  & 4 & $8.61 \pm 0.36$ \\  

 & period 3 & 1 & $7.60 \pm 0.12$ \\
 &  & 2 & $7.75 \pm 0.12$ \\  
 &  & 3 & $7.82 \pm 0.12$ \\  
 &  & 4 & $7.89 \pm 0.12$ \\    

 & period 4 & 1 & $8.21 \pm 0.30$ \\
 &  & 2 & $8.48 \pm 0.31$ \\  
 &  & 3 & $8.13 \pm 0.30$ \\  
 &  & 4 & $8.20 \pm 0.30$ \\ \cline{2-4}
 & Average & & $7.90 \pm 0.13$ \\  \bottomrule
\end{tabular}
\end{table}
 %%%%%%%%%%%%%%%%%%%%%%%%%%%%%%%%%%%%%%%%%%%%%%%%%%%%%%%%%%

%%%%%%%%%%%%%%%%%%%%%%%%%%%%%%%%%%%%%%%%%%%%%%%%%%%%%%%%%%
\subsection{Distribution of time duration between two errors}
\label{subsec:dt}
%%%%%%%%%%%%%%%%%%%%%%%%%%%%%%%%%%%%%%%%%%%%%%%%%%%%%%%%%%

The probability that SEE occurs is assumed to be constant. In that case, one
can show that the distribution of the time between two consecutive errors
follows a decreasing exponential 
\begin{equation}
p(\Delta t) \propto \exp(-\Delta t/\tau)
\end{equation}
where $\tau^{-1}$ corresponds directly to the probability. Looking at the $\Delta t$ allows then to verify the assumption
of a constant probability, and to extract this probability.

\begin{figure}[t]
 \begin{center}
   \includegraphics[width=1.0\textwidth]{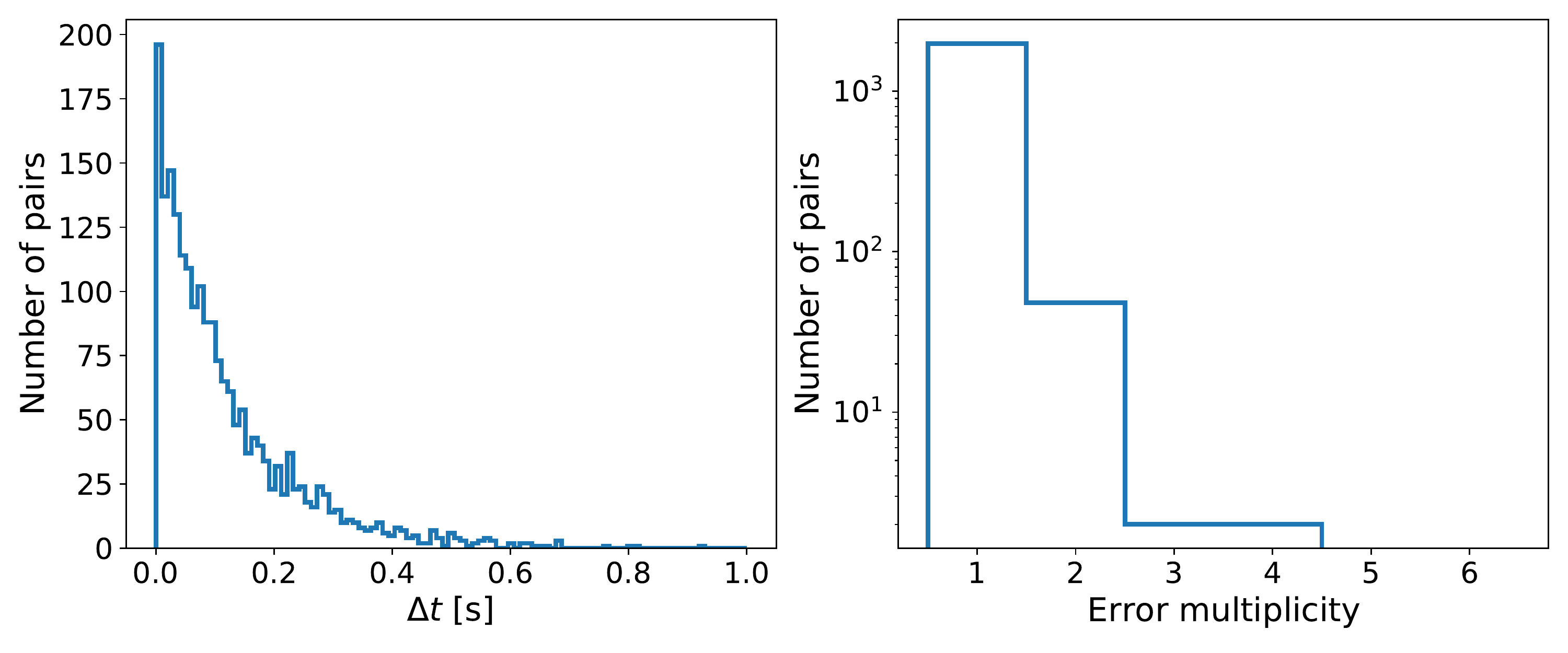}
   \caption{Time difference between two consecutive errors (left) and
     error multiplicity between two consecutive errors (right).
     \label{fig:deltat}}
  \end{center}
\end{figure}

In practice, error counters of every single bit are incremented every $25\,\nano\second$, but these individual counter values are summed to form one counter per group, every $\sim 8\,\milli\second$.
These values are read out and saved every $10\,\milli\second$. Therefore, the smallest measured $\Delta t$ value is $10\,$ms. A counter incrementation translates the number of errors, $n$, that occurred during this time window. When there is more than one error $n>1$ in a time window, it is assumed that those errors are equally distributed in time, meaning there are $n$ time intervals of $10/n\,$ms each.
The distribution of $n$ for all the $10\,$ms time windows, and the distribution of the time interval $\Delta t$ for which the counter incrementation is at least one are shown on Fig.~\ref{fig:deltat}.

\begin{figure}[t]
 \begin{center}
   \includegraphics[width=1.0\textwidth]{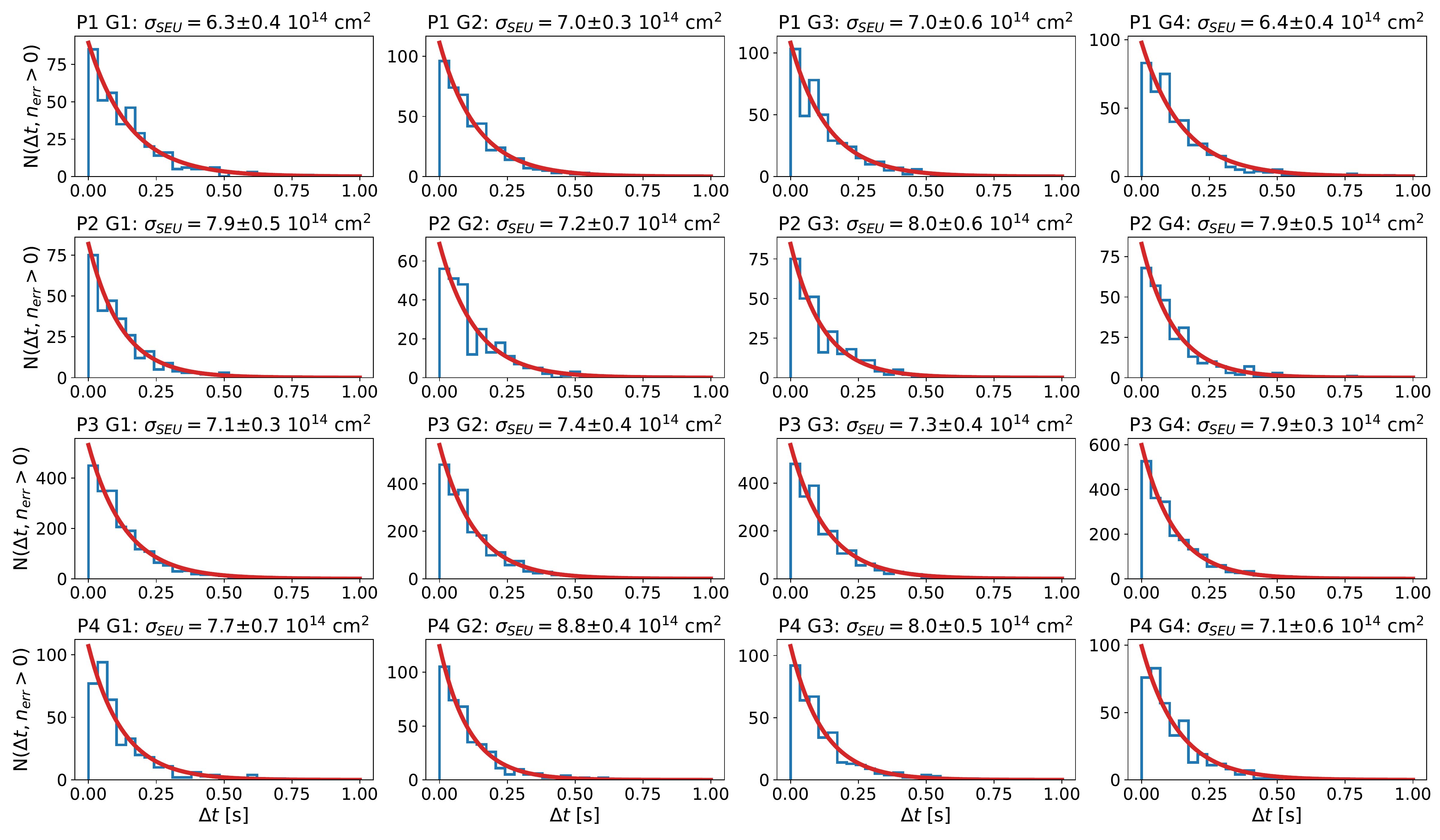}
   \caption{SEU cross-section computed from exponential fit of $\Delta t$ distribution for
     the four groups (line) of the four periods (column).
     The red line corresponds to the fitted exponential distribution. The index of the group
     and period as well as the fitted cross-section values are specified on top of each sub-plot.
    \label{fig:deltat_fit}}
  \end{center}
\end{figure}

The distribution and the fitted exponential function for all analyzed data are shown on Fig.~\ref{fig:deltat_fit}. As for Section~\ref{subsec:xsec},
the final value and uncertainty are taken from the average over the various datasets.
The obtained value of $\tau$ corresponds a SEU cross-section of $\sigma_{\mathrm{SEU}}^{1{\rightarrow}0}\,=\, (7.44 \pm 0.16)\, 10^{-14}\centi{\meter}^{2}$.
This value depends on the number of bins chosen to fit the $\Delta t$ distribution, increasing by $12\%$ when moving from 30 bins (used for the above value) to 100 bins.
As a consequence, this result is compatible with the measurement described in Section~\ref{subsec:xsec}.

%%%%%%%%%%%%%%%%%%%%%%%%%%%%%%%%%%%%%%%%%%%%%%%%%%%%%%%%%%
\subsection{Auto-correction failure due to time-overlapping SEU}
\label{subsec:computation}
%%%%%%%%%%%%%%%%%%%%%%%%%%%%%%%%%%%%%%%%%%%%%%%%%%%%%%%%%%

%%%%%%%%%%%%%%%%%%%%%%%% Table %%%%%%%%%%%%%%%%%%%%%%%%%%%
\begin{table}[t]
  \centering
  \caption{
    Predicted number of auto-correction failures due to uncorrelated SEUs, assuming an instantaneous fluence of    $10^{8}~\mathrm{n_{eq}^{1~\MeV}}\,\centi{\meter}^{-2}\,\second^{-1}$. $N_{\mathrm{HL-LHC}}^{\mathrm{trip~bit}}$
    is the expected number of auto-correction failures for one triplicated-bit.
    $N_{\mathrm{HL-LHC}}^{\mathrm{ASIC}}$ is the expected number of auto-correction failures for a 7200 triplicated-bits
    bits ASIC during 1 day.
\label{tab:failure}}
\begin{tabular}{ccc}
\toprule
& $N_{\mathrm{HL-LHC}}^{\mathrm{trip-bit}}$ & $N_{\mathrm{HL-LHC}}^{\mathrm{ASIC}}$ \\\midrule
$1\rightarrow 0$ & $(4.0 \pm 0.1)~10^{-13}$ & $(2.9 \pm 0.1)~10^{-9}$ \\\midrule
$0\rightarrow 1$ & $(3.0 \pm 0.4)~10^{-13}$ & $(2.1 \pm 0.3)~10^{-9}$ \\\bottomrule
\end{tabular}	
\end{table} 
%%%%%%%%%%%%%%%%%%%%%%%%%%%%%%%%%%%%%%%%%%%%%%%%%%%%%%%%%%

One mechanism that induces auto-correction failures is when two replicas of a single bit are affected by two SEUs during the correction cycle of $25~\nano{\second}$. The rate of the auto-correction failure can be deduced from the SEU rate.
The probability of two cell flips within a triplicated bit system can be written as:
\begin{equation}
P=C^{2}_{3}\int_{0}^{T_c}\sigma_{\mathrm{SEU}}\,f\,dt\,\int_{0}^{T_c}\,\sigma_{\mathrm{SEU}}\,f\,dt = 3 \sigma_{\mathrm{SEU}}^{2}\,f^{2}\,T_c^2
\label{eq:pdouble}
\end{equation}

where $T_c$ is equal to $25~\nano{\second}$ and $f$ the instantaneous fluence. 
Inner detectors in the ATLAS~\cite{atlas} and CMS~\cite{cms} experiments will be exposed to a typical integrated fluence of
$10^{15}\mathrm{n_{eq}^{1~\MeV}}$~\cite{tdrhgtd,tdritk} during the
HL-LHC. As indicated by equation~\eqref{eq:pdouble}, uncorrelated double flips depend quadratically on the intentaneous fluence.
The auto-correction failure rate is estimated assuming an instantaneous fluence of $10^{8}\mathrm{n_{eq}^{1~\MeV}}{\second}^{-1}$ and given in Table~\ref{tab:failure}. Each ASIC will have a negligible number of alterations of its registers during its lifetime due to uncorrelated double bit flips.

%%%%%%%%%%%%%%%%%%%%%%%%%%%%%%%%%%%%%%%%%%%%%%%%%%%%%%%%%%
\subsection{Observation of auto-correction failures}
%%%%%%%%%%%%%%%%%%%%%%%%%%%%%%%%%%%%%%%%%%%%%%%%%%%%%%%%%%
\label{sec:xsfail}
 The auto-correction failure rate was measured by seeking changes in the 64 register contents. Three measurements were performed with high intensity proton beam, using three register configurations:
 \begin{itemize}
 \item registers filled with {\ttfamily{1}},
 \item registers filled with a {\ttfamily{11111110}} frame,
 \item registers filled with {\ttfamily{0}}.
 \end{itemize}

\begin{figure}[t]
 \begin{center}
    \includegraphics[width=0.5\textwidth]{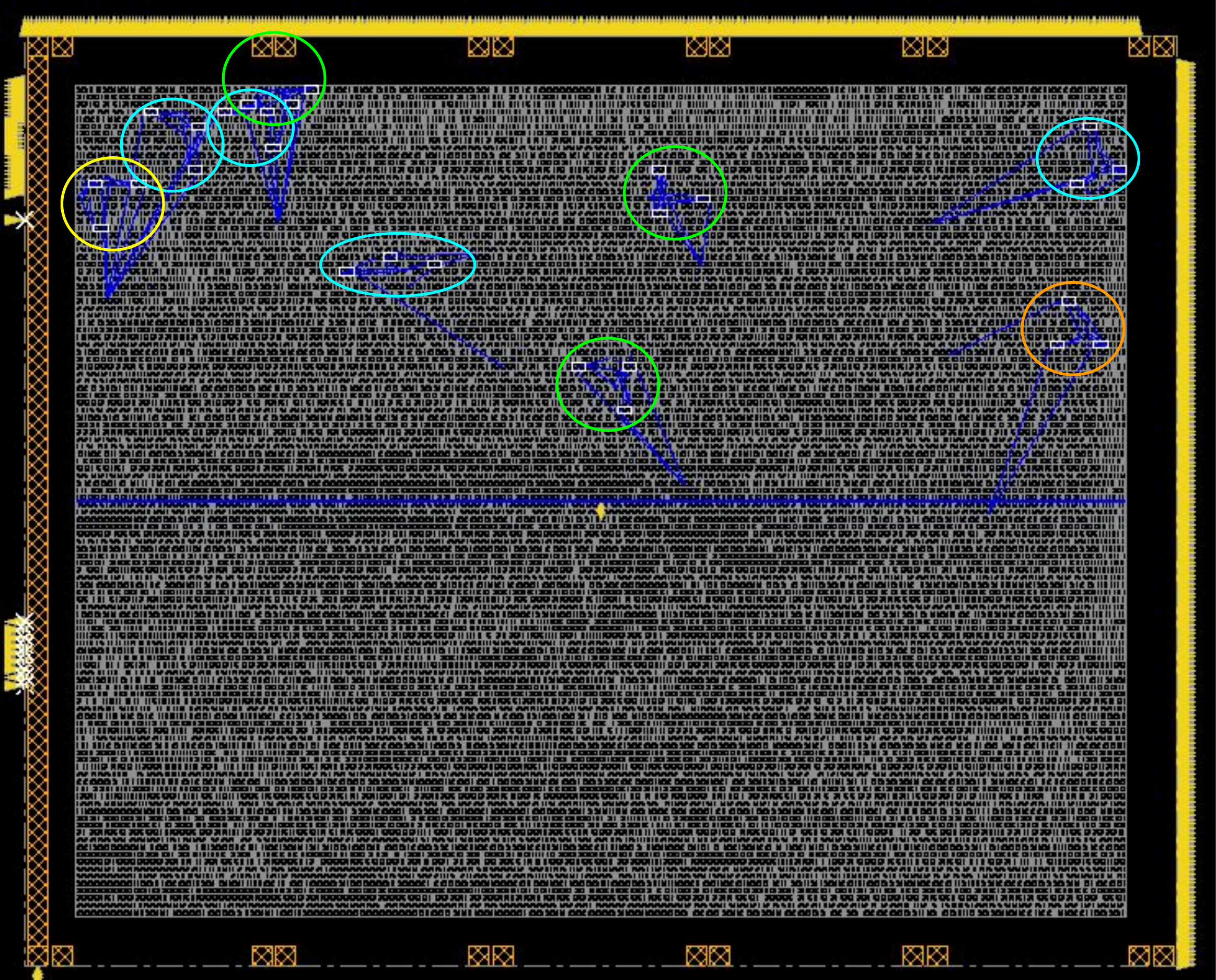}
  \caption{View of the ASIC where the flip-flops are highlighted by white rectangles. The 9 registers where a data corruption was observed are shown.\label{fig:failures}}
  \end{center}
\end{figure}

 The associated countings are provided in
 Table~\ref{tab:xsfail}. Significant statistical uncertainties are assigned to this measurement due to the small number of total errors. When no error is measured, a limit is set with a 95\% confidence level~\footnote{It can be shown that an upper limit of $\simeq{3}$ is assigned to the counting of 0 under a Poisson law with a confidence level of 95\%.}. It was checked that the failures affect single bits of different registers with no identified pattern. The location of these registers is shown on Fig.~\ref{fig:failures}.
The observed rate was extrapolated to HL-LHC integrated luminosity assuming a quadratic dependence to the fluence, as described by equation~\eqref{eq:pdouble}.
Comparing Table~\ref{tab:failure} and~\ref{tab:xsfail}, one can see that the observed  $0 \rightarrow 1$ uncorrected errors
are of the order of $10^6$ larger than what is expected from double SEUs.
This rate suggests that other mechanisms should be considered, including Single Event Transients (SET).
Additional measurements should be done in the future to confirm this rate.

 %The frequency at which uncorrected data corruption may occur was estimated. Based on the highest cross-section from
 %Table~\ref{tab:xsfail}, the frequency is about $10^{-8}\Hz$ per bit. 
%%%%%%%%%%%%%%%%%%%%%%%% Table %%%%%%%%%%%%%%%%%%%%%%%%%%%
\begin{table}[t]
  \centering
  \caption{
Number of auto-correction failures over all the registers, measured
using three data frame configurations. $N_{\mathrm{HL-LHC}}^{\mathrm{trip-bit}}$ ($N_{\mathrm{HL-LHC}}^{\mathrm{ASIC}}$)
is the expected number of failures per triplicated-bit (per 7200-triplicated bits ASIC) when extrapolating the observed
errors to the HL-LHC fluence.% The uncertainties are statistical.
\label{tab:xsfail}}
\begin{tabular}{ccccc}
\toprule
Frame & Duration [$\second$] & Errors & $N_{\mathrm{HL-LHC}}^{\mathrm{trip-bit}}$ & $N_{\mathrm{HL-LHC}}^{\mathrm{ASIC}}$\\\midrule
\ttfamily{11111111} & $610$ & $0$ & $<1.1~10^{-7}$ & $<0.8~10^{-3}$ \\\midrule
\ttfamily{11111110} & $740$ & $0$ & $<0.9~10^{-7}$ & $<0.7~10^{-3}$ \\\midrule
\ttfamily{00000000} & $530$ & $9$ & $(3.9 \pm 1.3)~10^{-7}$  & $(2.9\pm 0.9)~10^{-3}$ \\\bottomrule
\end{tabular}
\end{table} 
%%%%%%%%%%%%%%%%%%%%%%%%%%%%%%%%%%%%%%%%%%%%%%%%%%%%%%%%%%

%%%%%%%%%%%%%%%%%%%%%%%%%%%%%%%%%%%%%%%%%%%%%
\subsection{Cross-checks based on simulation}
\label{sec:sim}
%%%%%%%%%%%%%%%%%%%%%%%%%%%%%%%%%%%%%%%%%%%%%

The simulation randomly generates SEE events every $1\,$ms for each of the 1536 bits, with a given probability.
Every $8\,\milli\second$, the counters of individual bits are summed together to form the four general counters, as in the experimental data.
Every $10\,$ms, the cumulative error counts are written for each of these four counters.

\begin{figure}[t]
 \begin{center}
   \includegraphics[width=1.0\textwidth]{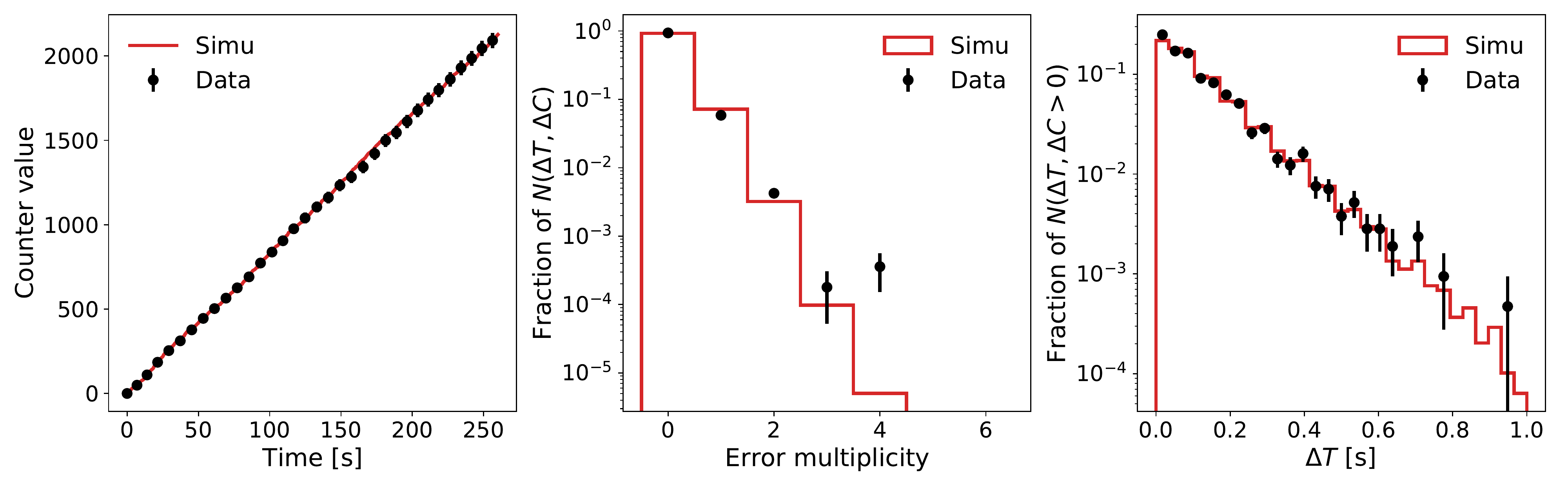}
   \caption{Comparison between data and simulation for the number of error as function time (left),
    error multiplicity between two consecutive errors (middle) and
     $\Delta t$ distribution (right).
    \label{fig:data_mc}}
  \end{center}
\end{figure}

Figure~\ref{fig:data_mc} shows a comparison between the observed data and the simulation for three observables, namely the number of errors as a function of time, the distribution of the time between two consecutive errors, and the error multiplicity between two consecutive errors. The assumption of a time-independent probability of error occurrence describes quite well the observed data. However, the simulation does not explain the error rate at a very short time since the observed number of 4 errors in $10\,$ms interval is much larger than the predicted one.
This observation goes in the same direction as the tension mentioned in Section~\ref{sec:xsfail}, where the number of explicit double errors measured every $25\,$ns is much higher than the double error rate computed from the single error probability.

 %% --
%% --------------------------------------

%% --------------------------------------
\section{Conclusion}

An ASIC prototype was designed to monitor the impact of radiation-induced errors on a $130~\nano{\meter}$ technology. It was exposed to a high-intensity proton beam.
Using specific counters, the SEU cross-section was measured using $62~\MeV$ protons considering both $1\rightarrow{0}$ and $0\rightarrow{1}$ bit flip observations. 
It was also shown that SEEs can be significantly mitigated by triple modular redundancy. While the single error rate obtained with a 130~nm TSMC technology follows
the expectations, the significant uncorrected error rate observed in the direct measurement requires further studies.
A new prototype with an improved implementation of the triplication using the latest TMRG tools will be designed and tested.
%The result has to be extrapolated at the scale of an entire device, considering the total number of bits and the expected fluence and energies.

%The implementation of the triplication can be improved using the latest TMRG tools. 

 %% --
%% --------------------------------------

%% --------------------------------------
\section*{Acknowledgements}
We would like to thank the MEDICYC staff from the Centre Antoine Lacassagne
for their availability and their 
precious help before and during the irradiation period. 
 %% --
%% --------------------------------------

\bibliographystyle{JHEP}
\bibliography{seu}

\providecommand{\href}[2]{#2}\begingroup\raggedright\begin{thebibliography}{10}

\bibitem{tmr}
J.~V. Neumann, \emph{{Probabilistic Logics}},  1956.

\bibitem{tmr2}
E.~Moore and C.~Shannon, \emph{Reliable circuits using less reliable relays},
  \href{https://doi.org/https://doi.org/10.1016/0016-0032(56)90559-2}{\emph{Journal
  of the Franklin Institute} {\bfseries 262} (1956) 191}.

\bibitem{method}
M.~Huhtinen and F.~Faccio, \emph{{Computational method to estimate single event
  upset rates in an accelerator environment}},
  \href{https://doi.org/10.1016/S0168-9002(00)00155-8}{\emph{Nucl. Instrum.
  Meth. A} {\bfseries 450} (2000) 155}.

\bibitem{dca}
G.~Balbi et~al., \emph{{Measurements of Single Event Upset in ATLAS IBL}},
  \href{https://doi.org/10.1088/1748-0221/15/06/P06023}{\emph{JINST} {\bfseries
  15} (2020) P06023} [\href{https://arxiv.org/abs/2004.14116}{{\ttfamily
  2004.14116}}].

\bibitem{medicyc}
P.~Mandrillon, F.~Farley, N.~Brassart, J.~Herault, A.~Susini and R.~Ostojic,
  \emph{{Commissioning and Implementation of the MEDICYC Cyclotron Programme}},
   in \emph{{12th International Conference on Cyclotrons and Their
  Applications}}, p.~E05, 1991.

\bibitem{medicyc2}
P.~Hofverberg, J.-M. Bergerot, R.~Trimaud and J.~Hérault, \emph{The
  development of a treatment control system for a passive scattering proton
  therapy installation},
  \href{https://doi.org/https://doi.org/10.1016/j.nima.2021.165264}{\emph{Nucl.
  Instrum. Meth. A} {\bfseries 1002} (2021) 165264}.

\bibitem{tmrg}
S.~Kulis, \emph{{Single Event Effects mitigation with TMRG tool}},
  \href{https://doi.org/10.1088/1748-0221/12/01/C01082}{\emph{JINST} {\bfseries
  12} (2017) C01082}.

\bibitem{seu.vs.energy}
R.~G. Alía et~al., \emph{{Single Event Effect cross section calibration and
  application to quasi-monoenergetic and spallation facilities}}, {\emph{EPJ
  Nuclear Sci. Technol. 4} {\bfseries 1} }.

\bibitem{atlas}
{\scshape ATLAS} collaboration, \emph{{The ATLAS Experiment at the CERN Large
  Hadron Collider}},
  \href{https://doi.org/10.1088/1748-0221/3/08/S08003}{\emph{JINST} {\bfseries
  3} (2008) S08003}.

\bibitem{cms}
{\scshape CMS} collaboration, \emph{{The CMS Experiment at the CERN LHC}},
  \href{https://doi.org/10.1088/1748-0221/3/08/S08004}{\emph{JINST} {\bfseries
  3} (2008) S08004}.

\bibitem{tdrhgtd}
{\scshape ATLAS Collaboration} collaboration, \emph{{Technical Design Report: A
  High-Granularity Timing Detector for the ATLAS Phase-II Upgrade}},  tech.
  rep., CERN, Geneva, Jun, 2020.

\bibitem{tdritk}
{\scshape ATLAS Collaboration} collaboration, \emph{{Technical Design Report
  for the ATLAS Inner Tracker Strip Detector}},  tech. rep., CERN, Geneva, Apr,
  2017.

\end{thebibliography}\endgroup

\end{document}